\documentclass{article}
\usepackage[cp1251]{inputenc}
\usepackage[dvips]{graphicx}
\begin{document}

\title{Spin-injection terahertz radiation in magnetic junctions}

\author{S.G. Chigarev$^1$, E.M. Epshtein$^1$\thanks{E-mail:
epshtein36@mail.ru}, Yu.V. Gulyaev$^1$, I.V. Malikov$^2$, \\
G.M. Mikhailov$^2$, A.I. Panas$^1$, V.G. Shofman$^1$, P.E. Zilberman$^1$\\ \\
$^1$\emph{V.A. Kotelnikov Institute of Radio Engineering and Electronics}\\
    \emph{of the Russian Academy of Sciences, 141190 Fryazino, Russia}\\ \\
    $^2$\emph{Institute of Microelectronics Technology and High Purity Materials} \\
    \emph{of the Russian Academy of Sciences, 142432 Chernogolovka, Russia}}
\date{}
\maketitle
\abstract{Electromagnetic radiation of 1--10 THz range has been found at
room temperature in a structure with a point contact between a
ferromagnetic rod and a thin ferromagnetic film under electric current of
high enough density. The radiation is due to nonequilibrium spin injection
between the structure components. By estimates, the injection can lead to
inverted population of the spin subbands. The radiation power exceeds by
orders of magnitude the thermal background (with the Joule heating taking
into account) and follows the current without inertia.}

\section{Experiment}\label{section1}
\begin{figure}
\includegraphics[width=90mm]{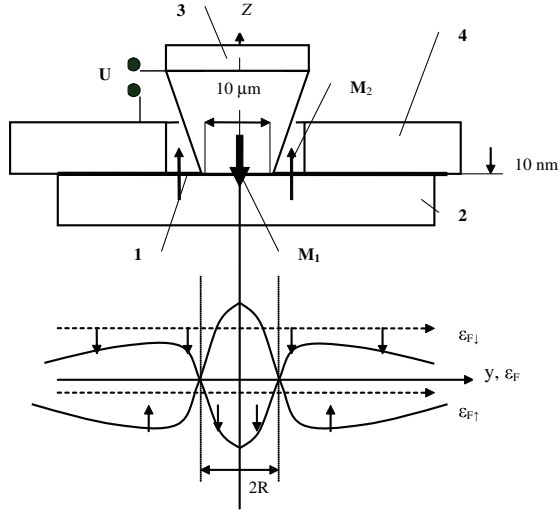}
\caption{The structure in study consists of a Permalloy (Py) film (1), a substrate
of oxidized silicon (2), a ferromagnetic rod (3), and a copper contact ring
(4). Voltage $U$ is applied between the rod and the ring, so that direct current
flows along the rod axis and through the Py film. The rod is magnetized along
its axis. The magnetization vector $\mathbf{M}_1$ inside the rod is shown with the heavy
arrow, the $\mathbf{M}_2$ component of the magnetization vector in the
closing magnetic field outside the rod is shown with two heavy arrows.
Note, that $\mathbf{M}_2$ depends on the properties of the magnetic
circuit under the substrate. The coordinate system is shown at the
bottom of the figure with $z$ axis coinciding with the rod axis. The bottoms of
the spin subbands in the Py film are presented (qualitatively) with solid lines,
the spin directions in the subbands are shown with thin arrows. The Fermi
level $\varepsilon_F$ splits to two quasilevels $\varepsilon_{F\downarrow}$
and $\varepsilon_{F\downarrow}$ (dashed lines) in presence of the current.}\label{fig1}
\end{figure}

The structure in study is an inhomogeneous magnetic junction. It consists
of two main ferromagnetic components: a hard magnetic rod of hardened
steel and a soft magnetic (Permalloy) thin film of nanosize thickness
(Fig.~\ref{fig1}). To simplify realization and calculations, the whole structure
has cylindrical form. The rod is sharpened at the end down to diameter of
$2R=10$--100 $\mu$m, the film thickness is $h=10$--100 nm.

The structure described has an important
property~\cite{Chigarev,Gulyaev1}, namely, because of the current
continuity, high spin-polarized current density may be reached in the film
exceeding by $2R/h\gg1$ times the current density in the rod. Then an
energy minimum for majority electrons appears in the rod center (Fig.~\ref{fig1}).
Such electrons concentrate near the minimum. With the current increasing,
the Fermi quasilevel $\varepsilon_{F\downarrow}$ rises and the electrons
from the minimum tend to penetrate into the film outside the rod. At the
distance $0<r-R<l$ from the rod edge smaller than the spin diffusion length
$l\sim20$--30\,nm, the majority spins are directed opposite to
$\mathbf{M}_2$, this means inverted spin population. The inversion is
promoted by an energy maximum for the electrons with minority spins which
appears near the rod axis (see Fig.~\ref{fig1}). The energy maximum leads to
withdrawal  of such electrons from the rod. Thus, spin separation occurs
in the junction because of its inhomogeneity, so that the electromagnetic
radiation power may be gained under intersubband transitions.

To detect radiation, we placed the receiver on the substrate side within
$L\ge100$\,mm distance of the substrate, so that the receiver could
rotate around the radiation source.

\begin{figure}
\includegraphics[width=100mm]{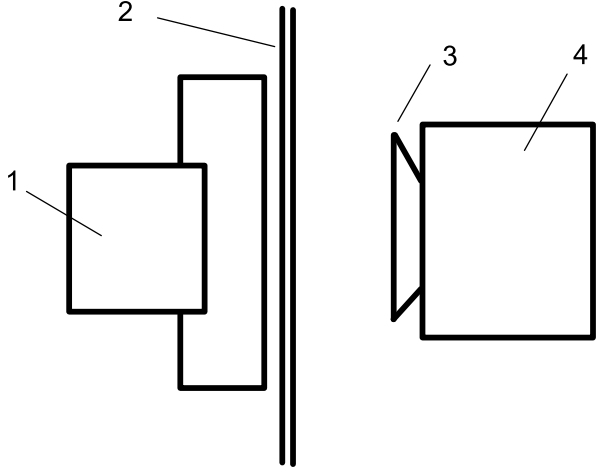}
\caption{The scheme of the receiving unit consisting of a radiator (1), a
low frequency filter (2), a high frequency filter (3), and Golay cell
(4).}\label{fig2}
\end{figure}

The problem of the current-driven spin
population in ferromagnetic films and observing radiation under
intersubband transitions evokes great interest for a long times.
Theoretical estimates of the expected THz frequencies have been made in a
number of works~\cite{Kadigrobov1,Kadigrobov2,Gulyaev2,Gulyaev3}. So, a Golay cell
was used as a detector and two filters to fix the frequency range, namely,
a metal grid with $125\times125\,\mu$m for low frequencies and a standard
TYDEX filter for high frequencies (Fig.~\ref{fig2}). As a result, the studied
frequency range was 1--10 THz. The receiver was in the wave zone,
because $L\gg\lambda$ condition was fulfilled, $\lambda$ being the
radiation wavelength.

\section{The measurement results and discussion}\label{section2}
\begin{figure}
\includegraphics[width=100mm]{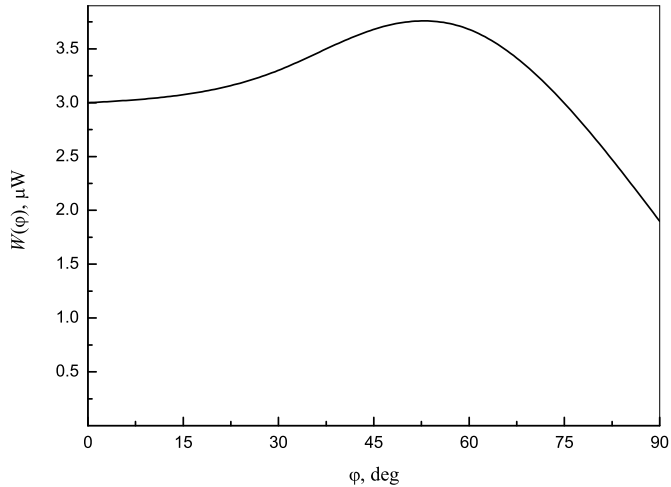}
\caption{The angular dependence of the received radiation power $W(\varphi)$
at current about $\sim400\,\rm mA$.}\label{fig3}
\end{figure}

Let us start with measuring dependence of the radiation intensity on the
observation angle $\varphi$, i.e., on the angle between the rod axis and
the normal to the objective plane. A typical angular dependence is shown
in Fig.~\ref{fig3}. It is seen that the radiation has no pronounced
directivity, so that an average power $\bar W\sim2.5\,\mu\rm W$ falls on
the objective at any angle. With the objective diameter of $\sim6\,\rm
mm$, this gives the total radiation power of $W_{\rm total}\sim10\,\rm mW$
to the complete solid angle $4\pi$. It is interesting to estimate the effective absolute
temperature $T$ of a body that radiates such a power. In the frequency
range in study, the Rayleigh--Jeans law is fulfilled,
\begin{equation}\label{1}
    W_{\rm total}=\frac{2\pi kTS}{c^2}\int_{f_1}^{f_2}f^2\,df,
\end{equation}
where $k$ is the Boltzmann constant, $c$ is the light velocity, $S$ is the
area of the radiating surface, $f$ is the radiation frequency. With
$S\sim0.1\,\rm{cm}^2$ and $f_1\sim10^{12}\,\rm Hz$, $f_2\sim10^{13}\,\rm
Hz$, we have $T\ge3000\,\rm K$. Such high temperatures were absent in the
experiment. Hence, the power of the receiving radiation is too high to be
explained in terms of the radiator heating.

\begin{figure}
\includegraphics[width=100mm]{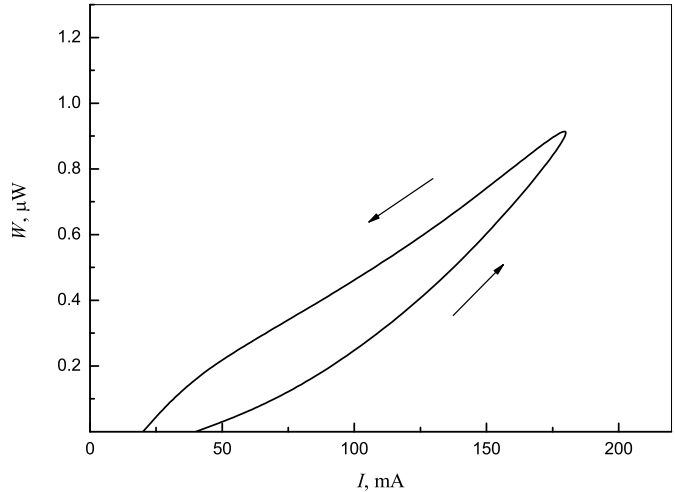}
\caption{The radiation power as a function of the current. The arrows
indicate direction of the current change.}\label{fig4}
\end{figure}

At the next stage, the receiving power has been measured as a function of
the current at $\varphi=0$. The results are shown in Fig.~\ref{fig4}.
The radiation appears starting with some threshold current. Under
decreasing the current, the backward branch of the $W(\varphi)$ dependence
does not coincide with the forward one, so that a loop appears. It may be
assumed that the sample properties change under current increasing,
so that we have somewhat another medium during the current decreasing.
Such a behavior is related with the Permalloy film coercivity. Indeed, the
magnetic flux from the rod reverses the film magnetization. The
injected conduction electrons contributes to this reversal, too. Under
current decreasing, the injection contribution decreases, but the film
state follows the current with some delay. Note, that the threshold
existence is an additional indication to non-thermal nature of the
radiation observed.

\begin{figure}
\includegraphics[width=120mm]{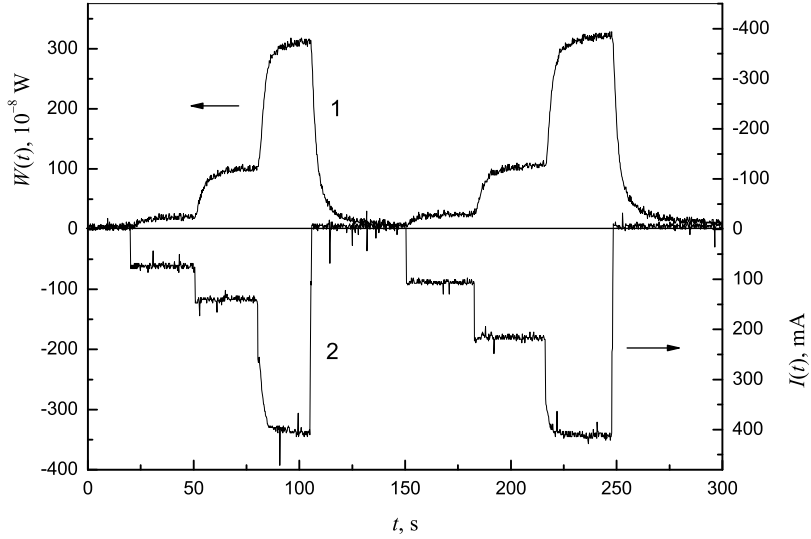}
\caption{The radiation power $W(t)$ (curve 1, the left ordinate axis)
and current (curve 2, the right ordinate axis) as functions of time.}\label{fig5}
\end{figure}

It is important to know, how the radiation follows the current under the
latter changing in time, with jumps and turning off. It is seen from
Fig.~\ref{fig5} that the radiation tracks all the variations of the
current. The only smooth character of the radiation pulse fronts under abrupt
jumps of the current and a small smooth ``tail'' near zero may be related
with thermal effects. Such a suppression of the thermal radiation is due
to using substrates with high thermal conductivity.
Early~\cite{Chigarev,Gulyaev1,Gulyaev4}, a fluoroplastic layer was used in
the substrate that slowed down the heat removal and led to intense heating
the sample by the current, so that the thermal radiation increased, while
the noh-thermal one suppressed. In present investigation, the sample
temperature remains low enough (no more than 60--100$^\circ$\,C), so that
the radiation of magnetic nature exceeding substantially the thermal background
becomes the main effect.

The spin-polarized current injects spin and disturbs spin equilibrium near
the contact between the junction layers (see review~\cite{Gulyaev5}).
According the estimates in Ref.~\cite{Gulyaev4}, this can lead to
inversion of the spin subband population. We show below that such an
inversion really takes place in our structure (Fig.~\ref{fig1}) under high
currents because of opposite directions of~$\mathbf{M}_1$ and
$\mathbf{M}_2$ magnetizations in the forward and closing magnetic fluxes.
The following equation is solved for the conduction electron spin
polarization $\mathbf P$ in the film~\cite{Gulyaev4}:
\begin{equation}\label{2}
  \frac{d^2\mathbf P}{dr^2}+(1-2\nu)\frac{1}{r}\frac{d\mathbf P}{dr}
  -\frac{\mathbf P-\bar{\mathbf P}}{l^2}=0,
\end{equation}
where $r$ is the distance from the rod axis, $\bar{\mathbf P}$ is the
equilbrium spin polarization,
\begin{equation}\label{3}
  \nu=\frac{R}{2l}\frac{j}{j_D},
\end{equation}
$j$ is the current density in the rod, $j_D=enl/\tau$ is a characteristic
current density ($j_D\sim10^9\,\rm A/cm^2$), $e$ is the electron charge,
$n$ is the electron density in the film, $\tau$ is the spin relaxation
time. Equation~(\ref{2}) is solved under condition that the polarization
is equilibrium ah large distance from the rod, i.e.,
$\mathbf{P}=\bar{\mathbf{P}}(\infty)$, while the spin current is
continuous at rod boundary $r=R$. Then we have at $r=R$~\cite{Gulyaev4}
\begin{equation}\label{4}
  \delta\mathbf R=\left(Q_1\hat{\mathbf{z}}\left(\hat{\mathbf{M}}_1
  \cdot\hat{\mathbf{M}}_2\right)-
  \bar{\mathbf{P}}\right)\frac{j(R)}{j_D}\frac{K_\nu(R/l)}{K_{\nu+1}(R/l)},
\end{equation}
where $\delta\mathbf P\equiv\mathbf P-\bar{\mathbf P}$ is the
nonequilibrium part of the spin polarization, $\hat{\mathbf z}$ is the
unit vector along the rod axis, $Q_1=\left|\sigma_\downarrow-\sigma_\uparrow\right|/
\left(\sigma_\downarrow+\sigma_\uparrow\right)$ is the spin polarization
of the rod conductivity, $\sigma_\downarrow$ and $\sigma_\uparrow$ are the
partial conductivities of the electrons with spin parallel and
antiparallel to $\hat{\mathbf{M}}_1$ vector, respectively, $\hat{\mathbf{M}}_{1,\,2}=
\mathbf{M}_{1,\,2}/\left|M_{1,\,2}\right|$, $K_\nu$ is the modified Bessel
function of the second kind. Because of $\left(\hat{\mathbf M}_1\cdot\hat{\mathbf
M}_2(R)\right)<0$, it follows from Eq.~(\ref{4}) that inverse population
can be reached, $\mathbf P(R)=-\hat{\mathbf z}\left|\mathbf
P(R)\right|$ at high enough $j/j_D$ ratio. The estimates in
Ref.~\cite{Gulyaev4} show that the inversion is reached at $j\sim10^8$--$10^9\,\rm A/cm^2$.

Note, that the dependence on the current in Eq.~(\ref{3}) appears not only
explicitly, but via the $\nu$ index of the Bessel function also. Because
of that fact, the $|P(R)|$ polarization and the radiation power can change
substantially (by more than 30\%). Such change may not be explained as
thermoelectric effects (Nernst--Ettingshausen effect and so on) which do
not exceed fractions of a percent in metals. The observed substantial
``nonreciprocity'' of the radiation may be additional argument in favor of
the spin injection mechanism.

Let us mention some check experiments which confirm magnetic nature of the
observed THz radiation. First, the non-thermal radiation was disappeared totally,
when the steel rod was replaced with a nonmagnetic (copper) one, the intensity was
decreased, and nonreciprocity was not observed.
Second, non-thermal radiation and nonreciprocity were disappeared also
under replacing the Permalloy film with molybdenum one, the steel rod being
retained. Third, attaching a core of soft magnetic steel made it possible
to carry the closing magnetic flux off the sample in study. As a result,
$\hat{\mathbf{M}}_2$ component was decreased down to zero. This manifested
in experiment as disappearance of THz radiation following the current, and
increasing in time thermal radiation due to nonradiative relaxation of the
spin polarized along $z$ axis.

\section{Summary}\label{section3}
We have observed rather intense radiation from a magnetic junction within 1--10 THz range. The
radiation power exceeds the thermal background by an order of magnitude.
The radiation has no pronounced directivity, its total power is about $\ge10\,\rm
mW$. The radiation appears with a current exceeding some threshold. The
radiation power rises with the current. Under decreasing current, the
radiation power falls and disappears at another threshold current. The
current polarity influence the radiation intensity changing it by more than
30\%.

As estimates show, an intense nonequilibrium spin injection occurs near
the contact in the structure under current flow. It is possible that
domains with inverse population of the spin subbands can appear. The spin
injection generates radiation with nonreciprocity relative to the current
polarity.

The authors are grateful to Yu.G. Kusraev, O.V. Betskii, G.A. Ovsyannikov
and Yu.G. Yaremenko for useful discussions.

The work is supported by the Russian Foundation for Basic Research, Grant
No. 02-00030.

\end{document}